\documentclass[11pt]{article}
\usepackage{epsfig}

\newcommand{\la}[1]{\label{#1}}
 
\newcommand{\be}{\begin{equation}}
\newcommand{\ee}{\end{equation}}
\newcommand{\ba}{\begin{eqnarray}}
\newcommand{\ea}{\end{eqnarray}}
\newcommand{\bi}{\begin{itemize}}
\newcommand{\ei}{\end{itemize}}
\newcommand{\rmi}[1]{{\mbox{\scriptsize #1}}}
\newcommand{\nr}[1]{(\ref{#1})}
\newcommand{\tr}{{\rm Tr\,}}

\newcommand{\nn}{\nonumber \\}
\newcommand{\fr}[2]{{\frac{#1}{#2}}}
\newcommand{\msbar}{\overline{\mbox{\rm MS}}}

\renewcommand{\(}{\left(}
\renewcommand{\)}{\right)}

\newcommand{\lk}{\left[}
\newcommand{\rk}{\right]}
\newcommand{\ld}{\left.}
\newcommand{\rd}{\right.}

\newcommand{\bmu}{\bar{\mu}}

\newcommand{\RR}{{\rm I\kern -.2em  R}} 
\newcommand{\eq}{Eq.~}
\newcommand{\eqs}{Eqs.~}
\newcommand{\fig}{Fig.~}

\newcommand{\se}{Sec.~}

\def\lsi{\raise0.3ex\hbox{$<$\kern-0.75em\raise-1.1ex\hbox{$\sim$}}}
\def\gsi{\raise0.3ex\hbox{$>$\kern-0.75em\raise-1.1ex\hbox{$\sim$}}}
\newcommand{\lsim}{\mathop{\lsi}}
\newcommand{\gsim}{\mathop{\gsi}}

\makeatletter \@addtoreset{equation}{section} \makeatother

\makeatletter
\renewcommand\section{\@startsection {section}{1}{\z@}%
                                   {-5.5ex \@plus -1ex \@minus -.2ex}
                                   {2.3ex \@plus.2ex}%
                                   {\normalfont\large\bfseries}}
\renewcommand\subsection{\@startsection{subsection}{2}{\z@}%
                                     {-3.25ex\@plus -1ex \@minus -.2ex}%
                                     {1.5ex \@plus .2ex}%
                                     {\normalfont\normalsize\bfseries}}
\renewcommand\thesection {\@arabic\c@section}
\renewcommand\thesubsection   {\thesection.\@arabic\c@subsection}
\renewcommand{\@seccntformat}[1]{%
\csname the#1\endcsname.\hspace{1.0em}}
\makeatother

\begin{document}
 
\begin{titlepage}
\begin{flushright}
BI-TP 2004/09\\
CERN-PH-TH/2004-055\\
EPFL-ITP-LPPC/2004-1\\
HIP-2004-14/TH\\
hep-ph/0404058\\
\end{flushright}
\begin{centering}
\vfill
 
\mbox{\Large{\bf Effective gauge theories on domain walls
via bulk confinement?}}

\vspace{0.8cm}
 
M. Laine$^{\rm a}$, 
H.B. Meyer$^{\rm b}$, 
K. Rummukainen$^{\rm c,d,e}$, 
M. Shaposhnikov$^{\rm f}$ 

\vspace{0.8cm}

{\em $^{\rm a}$%
Faculty of Physics, University of Bielefeld, 
D-33501 Bielefeld, Germany\\ }

\vspace{0.3cm}

{\em $^{\rm b}$%
Theoretical Physics, University of Oxford, 
1 Keble Road, Oxford, OX1 3NP, UK\\ } 

\vspace{0.3cm}

{\em $^{\rm c}$%
Department of Physics, Theory Division, CERN, CH-1211 Geneva 23,
Switzerland\\ }

\vspace{0.3cm}

{\em $^{\rm d}$%
Department of Physics, University of Oulu, 
P.O.Box 3000, FIN-90014 Oulu, Finland\\ }

\vspace{0.3cm}

{\em $^{\rm e}$%
Helsinki Institute of Physics,
P.O.Box 64, FIN-00014 University of Helsinki, Finland\\}

\vspace{0.3cm}

{\em $^{\rm f}$%
Institute of Theoretical Physics, 
Swiss Federal Institute of Technology (EPFL), 
BSP-Dorigny, CH-1015 Lausanne, Switzerland }

\vspace*{0.8cm}
 
\end{centering}
 
\noindent 
We study with lattice techniques the localisation of gauge fields on
domain wall defects in 2+1 dimensions, following a scenario
originally proposed by Dvali and Shifman for 3+1 dimensions, based on
confining dynamics in the bulk. We find that a localised gauge
zero-mode does exist, if the domain wall is wide enough compared with
the confinement scale in the bulk. The range of applicability
of the corresponding low-energy effective theory is determined by the
mass gap to the higher modes. For a wide domain wall, this mass gap
is set by ``Kaluza--Klein modes'' as determined by the width. It is
pointed out that in this regime the dynamical energy scales generated
by the interactions of the localised zero-modes are in fact higher
than the mass gap. Therefore, at least in 2+1 dimensions,  the
zero-modes alone do not form a low-energy effective gauge theory of a
traditional type. Finally, we discuss how the situation is expected
to change in going to 3+1 dimensions.
\vfill
\noindent
 

\vspace*{1cm}
 
\noindent
April 2004 

\vfill

\end{titlepage}

\setcounter{footnote}{0}

\section{Introduction}
\la{se:intro}

In brane-world scenarios with infinite \cite{Rubakov:bb,Akama:jy,rs}
or large \cite{add} extra dimensions, it is assumed that all the
fields of the Standard Model have wave functions that are localised
in transverse directions, making the physics to be four-dimensional
(4d) at small energy scales.  The field-theoretical realisation of
branes is related to topological defects --- stable solutions of the
classical equations of motion, which depend on the transverse
coordinates only.  In five dimensions, the necessary defect  is a
domain wall, in six dimensions it is a (4d)  string, in seven
dimensions a monopole, etc. (For an explicit construction  of these
solutions in the case of localisation of gravity, for instance,   
see Refs.~\cite{DeWolfe:1999cp,Giovannini:2001hh,Roessl:2002rv}.)

Small perturbations of the fields around the topological defect
solutions may form  a low-dimensional effective theory ---
potentially the Standard Model (for a general discussion of
perturbations see, e.g.,  Ref.~\cite{Randjbar-Daemi:2002pq}). Apart
from the requirement that the wave function of the ``fundamental''
(or lowest-energy) mode  be centered around the brane in transverse
directions,   such that it has long-wavelength perturbations in four 
directions only, a successful localisation poses other constraints as
well. Indeed, either the perturbations of all the higher modes should
be separated from those of the fundamental mode by a sufficient mass
gap,   in order not to be observable in low-energy experiments, or,  
if there is no mass gap, the perturbations of the higher modes should
interact very  weakly with those of the fundamental one.

The possible solutions to these requirements can roughly be divided
into two classes (for a recent review, see Ref.~\cite{vr}).  On one
hand there are mechanisms which, in one way or the other, invoke
effects related to gravity (such as a ``warped'' metric
\cite{Rubakov:1983bz,rs}). In this way  scalars
\cite{baj,Oda:2000zc},  fermions \cite{baj}, 
\cite{Randjbar-Daemi:2000cr}--
\cite{Randjbar-Daemi:2003qd},  Abelian gauge fields
\cite{Oda:2000zc,Kehagias}, 
\cite{Randjbar-Daemi:2003qd}--
\cite{Giovannini:2002sb}  and gravity \cite{rs}  can be localised  on a
brane, although the construction of the full Standard Model is still
far from being achieved. On the other hand, there  are mechanisms
which work in flat spacetime and are purely field theoretic  in
origin. For example, the localisation of fermions on a brane may be
due to the presence of fermionic zero-modes on topological
defects~\cite{Rubakov:bb,dbk}, while scalars can be attached to the
brane through their interactions with  the field forming the
topological defect (and the perturbations of the defect forming field
itself constitute a light scalar field).

The most difficult problem along these lines seems to be the
localisation of massless non-Abelian gauge fields (for a general
discussion,  see Ref.~\cite{Dubovsky:2001pe}). If the mechanism
related to gravity is used, for instance, then  for typical warp
factors, the bulk gauge fields have a spectrum which is not separated
by a mass gap from  the localised modes, so that no effective field
theory can be constructed.\footnote{%
  Both massless and  massive
  vector fields {\em can}  be localised on a brane,  such that the
  fundamental mode {\em is} separated from the higher ones by a
  non-zero mass gap, if the warp factor is tuned  appropriately ``by
  hand'' \cite{Shaposhnikov:2001nz,us} (see also Ref.~\cite{dfkk}).} 
At the same time, in the case without gravity,  no perturbative
mechanism of gauge field localisation is available at present,  as
far as we know.\footnote{%
  Although of significant interest, we do
  not consider  mechanisms related to a high degree of unbroken
  supersymmetry in  this paper (see, e.g., Ref.~\cite{sy} and
  references therein).}

A very interesting non-perturbative purely field theoretic idea for
gauge field localisation was put forward some time ago by Dvali and
Shifman~\cite{ds}. It uses the fact that non-Abelian gauge theories
are strongly coupled in the infrared  and that a mass gap can be
generated for the bulk modes by confinement effects. In short:
consider a confining theory in the bulk, based on some group G.
Construct a topological defect that ``eats up" the necessary number
of dimensions and breaks this symmetry down to G$'$ inside the
defect, while leaves it intact outside. Then the gauge fields related
to G$'$ are localised on the defect and are separated by a mass gap
from the bulk modes, which are massive because of confinement. The
resulting low-energy effective  theory is thus a four-dimensional
gauge theory based on G$'$.

Of course, to use this idea for the construction of a realistic
theory, one would first have to find a confining gauge theory in $4 +
n$ dimensions, with $n \ge 1$.  Non-Abelian gauge theories of the
usual type are, however,  not renormalisable when
extra dimensions are involved.  In five dimensions, for instance,
lattice simulations \cite{mc} do not reveal any second order phase
transition that can be used for a continuum formulation of a
confining theory.\footnote{%
 For recent discussions and references see, e.g., Refs.~\cite{gies}.}
Thus, to have a simple renormalisable framework,
Dvali and Shifman  assumed the bulk dynamics to be that of 4d gauge
theory, so that the low-energy dynamics is that of three-dimensional
(3d) gauge theory. 

Even if plausible after this simplification,  the idea does still
involve some untested assumptions. First of all, it may not be
immediately clear why the non-perturbative confinement effects, 
acting in the bulk and admittedly suppressed on the brane, could not
generate an effective mass term for the brane gauge fields; this
depends after all on the precise boundary conditions that the bulk
phase  poses on the localised modes~\cite{ahs}. Another question,
necessary for understanding whether the effective theory is truly
lower dimensional than the bulk theory,  is related to the magnitudes
of the confinement induced mass gap $M_c$, the typical energy scale
of the low-dimensional theory $m_l$, and the masses of higher
``Kaluza-Klein'' (KK) excitations $M_k$, localised on  the
brane.\footnote{%
  Even though there is no compact dimension involved, 
  we refer to the tower of localised states above the 
  fundamental mode as Kaluza-Klein 
  excitations, due to the fact that they have a similar 
  spectrum.} 
To hide extra dimensions at small energies, the scale $m_l$ must
be much smaller than the masses of particles in the bulk as well as
those of the KK-excitations: $ m_l \ll M_k,~m_l \ll M_c$. Otherwise,
the effects of higher dimensions are not suppressed.  This point has
not been discussed in Refs.~\cite{ds,ahs}, as far as we can judge. 

Given the open ends,  it is the purpose of this paper to test the
idea of Ref.~\cite{ds}  with simple non-perturbative lattice
simulations. We consider SU(2) gauge theory coupled to an adjoint
scalar field with an effective space-dependent mass parameter induced
by the  kink solution, as in the original proposal~\cite{ds}. Since
going to higher dimensions leads us to a shaky ground we will deal,
following Ref. \cite{ds}, with renormalisable theories only, i.e.
consider at most a 4d bulk.  For better numerical resolution and
since the physics arguments are almost unchanged (see below),   we
will however reduce the dimensionality to be three in total,  such
that the low-energy effective theory is supposed to be
two-dimensional (2d).  

Our lattice simulations do support the Dvali-Shifman conjecture on
the existence of a gauge field localised on the brane. In other
words, if the dynamics of the theory is probed with external 
sources separated by large distances along a (wide enough) brane, 
a force specific to a massless localised gauge mode 
will indeed manifest itself. 
We find, however, that in this limit there is no hierarchy of
(dynamical) scales between the 2d localised mode and certain
higher excitations and, therefore, that the low-energy theory 
cannot be considered to be a traditional 2d gauge theory.

This paper is organised as follows.  In \se\ref{se:overview} we
review the basic mechanism as proposed by Dvali and
Shifman~\cite{ds}.  In \se\ref{infvol} we derive its signatures for a
specific  observable, the static force between two heavy test charges
living on the  brane, and in~\se\ref{finvol} discuss how the
signatures change if the  world-volume is finite.   In
\se\ref{se:latt} the system is put on the lattice and the observables
are written in a form accessible to Monte Carlo simulations.  The
main results are presented in \se\ref{se:results},  and our
conclusions as well as a brief outlook, in \se\ref{se:concl}.

\section{Overview of the Dvali-Shifman mechanism}
\la{se:overview}

The model suggested by Dvali and Shifman has two scalar fields, 
a gauge singlet $\eta$ and an adjoint scalar $\chi$, and has the
following Euclidean action in the case of a 3d bulk:
\ba
 S_\rmi{E} 
  &\equiv&   \int  {\rm d}^3 x\, {\cal L}_\rmi{E} 
 \;, \\
 {\cal L}_\rmi{E} 
  &\equiv & \fr12 \tr F_{kl}^2 + \tr [D_k,\chi]^2 + 
 \lambda\,\tr [\chi^2]^2 +\gamma\, \tr [\chi^2]\left({u}^2-
 {v}^2+\eta^2\right) \nn 
 &+&\frac{1}{2}(\partial_\mu\eta)^2+ \frac{1}{4}
 \kappa \left(\eta^2-{v}^2\right)^2, 
 \la{lfull}
\ea
where 
$k,l=1,...,3$,
$D_k = \partial_k + i g A_k$, 
$A_k = A_k^a T^a$, $\chi = \chi^a T^a$, $F_{kl} = (1/ig) [D_k,D_l]$, 
and $T^a$ are the Hermitean generators of SU($N_c$), 
normalised as $\tr [T^a T^b] = \delta^{ab}/2$.  Summation over repeated 
indices is understood. The coupling constants $\lambda, \gamma, \kappa$ 
and the parameters ${u}^2$ and ${v}^2$ are assumed to be positive, 
and $v^2 > u^2$.

The classical vacuum of the theory is at $\chi=0$, $\eta = \pm
{v}$ and the perturbative spectrum consists of the scalar singlet
with the mass  
$m_\eta^2= 2 \kappa {v}^2$,   
the scalar triplet with the mass
$m_\chi^2=\gamma {u}^2$ and massless gauge bosons corresponding to
the gauge group SU($N_c$). Because of strong coupling in the
infrared, the vector boson spectrum aquires a mass gap of the order
of the 3d confinement scale, $\Lambda \sim g^2$. In other words, 
gluons form bound states --- glueballs --- with a mass of the order
of $\Lambda$.  Depending on the relation between $\Lambda$ and
$m_\chi$, bound states of the triplet scalar and gluons can have 
masses $\mathcal{O}(\Lambda)$ 
or $\mathcal{O}(m_\chi)$. 

Now, the model of \eq\nr{lfull} always has  a kink solution, $\chi=0$, 
$\eta(z)= v \tanh(\frac{m_\eta z}{2})$.  
By inspecting whether the fluctuation Hamiltonian 
around this solution develops a negative eigenvalue, it is seen that 
the solution is unstable against $\chi$-field condensation near $z=0$, 
provided that
\be
 \gamma {v}^2> m_\chi^2 + \frac{1}{2} m_\chi m_\eta
 \;. \label{cond}
\ee
In this case the stable classical solution contains a non-zero $\chi$
field as well, and has the asymptotics $\chi_3 \to 0$, $\eta \to \pm
{v}$ at $z \to \pm \infty$, whereas the components $\chi_1$ and
$\chi_2$ can be chosen to be zero. For a general choice of parameters
the explicit solution can easily be constructed numerically. The
analytic form can be found for a specific choice of parameters,
namely for
\be
 2 \gamma {u}^2=\frac{\kappa \lambda - \gamma^2}{\lambda-\gamma} {v}^2
 \;, 
\ee
and is given by
\be
 \eta(z)={v} \tanh (m_\chi z)\;,~~
 \chi_3(z)=\sqrt{\frac{\kappa-\gamma}{\gamma-\lambda}}
 \frac{{v}}{\cosh(m_\chi z)}
 \;.
\ee
In the special case $\kappa=\gamma=\lambda$, 
$\eta(z)$ remains the same while
\be
 \chi_3(z)=\frac{\sqrt{v^2-2u^2}}{\cosh(m_\chi z)}
 \;.
\ee

To get the mechanism to work, the following choice of parameters is
proposed. Inside the defect the SU($N_c$) symmetry is partially
broken to G$'$, and the masses of the corresponding vector bosons are
assumed to be large compared with the confinement scale $\Lambda$, 
\be
 \frac{\gamma({v}^2-{u}^2)}{\lambda} \gg g^2 
 \;. \la{largemW}
\ee
At the same time, the width of the domain wall is chosen to be larger
than the inverse of the confinement scale, $g^2 \gg m_\chi$.  These two
requirements suppress the non-perturbative confinement  effects
inside the defect. Together with~\eq\nr{cond},  they pose
restrictions on the parameter space, but  given that there are
several parameters at our disposal  ($u,v,\lambda,\gamma,\kappa$),
all the requirements can  easily be satisfied simultaneously. 

At the quadratic level the spectrum of perturbations around the
domain wall contains a few normalizable localised scalar modes (there
is  one zero-mode associated with brane translations and
fluctuations), a continuous spectrum of scalar excitations that
starts from $\min(m_\eta, m_\chi)$, and a gapless continuous spectrum
of vector excitations corresponding to the unbroken group G$'$. 
Now,  because the true non-perturbative spectrum of bulk gauge
excitations is massive, it was conjectured in Ref.~\cite{ds} that the
vector bosons of the unbroken group cannot escape the brane and thus
the true spectrum of gauge excitations around the domain wall does
contain  normalisable vector modes related to G$'$. If true, the
low-energy effective theory is just a 2d gauge theory in our case.

Further support for the Dvali-Shifman idea has been provided by 
Arkani-Hamed and Schmaltz, who argued~\cite{ahs},  based on the 't
Hooft -- Mandelstam picture of confinement, that the bulk acts as a
Neumann boundary  condition for the gauge fields corresponding to
G$'$,  such that there is a true localised  zero-mode, unlike in the 
case of a Dirichlet boundary condition, whereby a  zero-mode is
excluded.\footnote{%
   For further work on the topic see, e.g., Ref.~\cite{nt}.}  
Since the physics of confinement is involved, however, a
non-perturbative check of the mechanism would be welcome, and this is
one of the aims of the present work. Another problem, related to 
mass gaps and the dimensionality of the low-energy theory, has
been mentioned already in the Introduction, and will be explained in
more detail below.

%
\section{How to probe the properties of the low-energy effective theory?}
\la{infvol}

Since our main interest here is in the gauge fields, we will ignore
in the following all the dynamics related to the scalar singlet field
$\eta$. To achieve this we go to the rest frame of the domain wall,
and also treat it as infinitely rigid, which means that we remove the
3d zero-mode related to translations and fluctuations of the domain
wall.

The remaining dynamical degrees of freedom constitute 
the SU($N_c$) gauge + adjoint Higgs theory, 
formally  defined by the action
\ba
 S_\rmi{E} 
 & \equiv &  \int  {\rm d}^3 x\, {\cal L}_\rmi{E} 
 \;, \\
 {\cal L}_\rmi{E} 
 & \equiv & \fr12 \tr F_{kl}^2 + \tr [D_k,\chi]^2 + 
 m^2(z)\tr [\chi^2] +\lambda (\tr [\chi^2])^2, 
 \la{leff}
\ea
where $m^2(z)$ is simply some profile for the mass of $\chi$, which
we fix ''by hand'' to be
\be
 m^2(z) \equiv m_2^2 + \frac{m_1^2 - m_2^2}{\cosh^2 (z/\ell)} 
 \;, \la{m2z}
\ee
where, in the notation of \se\ref{se:overview}, 
$m_2^2 \equiv m_\chi^2 = \gamma {u}^2$ 
represents the mass of the scalar triplet outside the brane, while
$m_1^2 = \gamma ({u}^2 - {v}^2)$ is negative and is related to the
mass of the scalar boson inside the brane. As already discussed (cf.\
\eq\nr{largemW}),  we assume that the masses of the vector bosons
inside the brane are large,  $m_W^2 \sim - g^2 m_1^2/\lambda \gg
g^4$. The domain wall width, $\ell$, is for generality now treated as
a parameter independent of $m_\chi$, and is supposed to be large
enough, 
\be
 \ell \gg \frac{1}{g^2}
 \;, \label{as}
\ee
to suppress the influence of bulk confinement on the localised gauge
field. In addition, to have condensation of $\chi$ inside the brane,
we must require 
(cf.\ \eq(\ref{cond})) that $m_1^2 \lsim - m_2/\ell$.

The dimensionless combinations of the parameters, determining the
actual dynamics,  can be chosen as
\be
 \alpha \equiv \ell g^2 \;, \quad
 x \equiv \frac{\lambda}{g^2} \;, \quad
 y_{1,2} \equiv \frac{m^2_{1,2}}{g^4} 
 \;, \la{alpha} 
\ee
where the mass parameters (the only ones requiring renormalisation in
three dimensions) are for convenience assumed to be evaluated in the
$\msbar$ scheme at the scale $\bmu = g^2$.

Let us assume now that the Dvali-Shifman conjecture is correct and
estimate the parameters of the localised gauge theory. Since the wave
function of the massless mode is localised on a length of the order
of the width of the domain wall, the 2d effective gauge coupling
$g_2$ is simply 
\be 
\frac{1}{g_2^2} \sim \frac{\ell}{g^2}~.
\label{2dg}
\ee
If two opposite static test charges are put on the brane, one would
then expect that they are attracted, at large distances  (to be
specified presently) with a force $F \sim g_2^2$ that does not depend
on the distance. Note that this holds also for G$'$ = U(1),  since
the Coulomb potential is linear in 1+1 dimensions.

Besides the massless mode one would expect to  also have a whole tower
of states with the same quantum numbers and with an energy spacing 
of the order of $E \sim 1/\ell$, with $E$ coming simply from the
uncertainty principle.  These states are not seen  at distances $r
\gg \ell$ or at energies $E \ll 1/\ell$, so that the force derived
above should be valid for $r \gg \ell$. For $ m_W^{-1} \ll r \ll
\ell$ the  2d Coulomb law is expected to be replaced by the 3d
Coulomb law with the force $F \sim {g^2}/{r}$, while at even smaller
distances the massive $W$ will contribute as well,  changing the
numerical coefficient in front  of the $1/r$-dependence of the
force. 

If, on the contrary, the localised zero-mode acquires a mass
$m_\gamma$ due to interactions with the bulk modes, the force between
the test charges at $r \gg \ell$ will have a Yukawa character, $F
\sim g_2^2 \exp(- m_\gamma r)$, whereas for smaller  $r$ the
behaviour is still as described above.

These two different behaviours of the force at $r \gg \ell$ can be
distinguished in lattice simulations, as will be described below, so
that a conclusion can be reached on the existence of a localised
vector zero-mode. As we will see, the presence of a vector zero-mode
is confirmed, and the dependence of the 2d coupling constant on the
width of the domain wall (\eq\nr{2dg}) is also  found to be as
expected.

Let us now discuss whether the low-energy theory can indeed be
considered to be 1+1 dimensional electrodynamics. If so, the typical
energy scale $E_0$ of the 1+1 dimensional theory must be smaller than
the mass gap to the first excited KK mode. To be concrete, let us
imagine adding fermions to the 3d theory and let them interact with
the singlet field in a way that ensures the existence of fermionic
zero-modes. The gauge coupling of these fermions is $g_2$. Then, the
theory of the zero-modes (fermionic and vector) is simply the
Schwinger model, the solution of which tells that bosonic scalar
states with the mass $E_0 \sim g_2$ are formed. The KK tower of
states decouples, provided that $E_0 \ll 1/\ell$. This inequality,
together with \eq(\ref{2dg}) and the estimate $E_0\sim g_2$, gives
$\ell \ll g^2$, which is in contradiction with the initial
assumption of \eq(\ref{as}). In other words, we do not expect the
low-energy effective theory to be the Schwinger model, it
will rather be a more
complicated 1+1 dimensional theory incorporating not
only the massless mode but also its KK excitations. 

This expectation can also be tested on the lattice. Indeed, if for
some reason the mass gap to the KK excitations is considerably higher
than $1/\ell$ and corresponds to some distance scale $\ell_0 \ll
\ell$, then the 1+1 dimensional Coulomb law will be valid to
considerably smaller distances. As we will see in the lattice
simulations, the deviations start really from $r \sim \ell$, and are
characterised by a mass gap $\sim 1/\ell$.  Therefore, the low-energy
theory, though giving an apparently 1+1 dimensional  force at the
distance scales $\gg \ell$, is in general not that of the zero-modes
alone.

We also note that the other logical possibility, $\ell < {1}/{g^2}$,
does not lead to 1+1 dimensional electrodynamics either. In this
case the spread of the potentially massless localised mode is given
by the inverse confinement scale, $1/g^2$ in our case. Thus, the
effective 2d gauge coupling is $g_2^2 \sim g^4$, and, therefore, the
typical 2d energy scale is of the same order as the mass gap to the
bulk modes.

To conclude this section, we note that the specific line of reasoning
above is related to the case of 3$\to$2 compactification. In the last
section we however provide arguments that the pattern is essentially
the same also for  4$\to$3 compactification, in the generalised case
that G$'$ is a non-Abelian group.  Unfortunately, we have nothing to
say about (the most interesting) higher dimensional case. 

%
\section{Finite-size scaling of the Abelian static force in a periodic box}
\la{finvol}

As explained in the previous section,  the probe to be used in order
to test the conjecture, is the static force between infinitely heavy
test charges living on the brane.  The motivation for this choice is
that theoretical predictions, as reviewed in the previous section,
are unambiguous, and that lattice measurements, as discussed in
\se\ref{se:latt}, can be made rather precise, employing recent
technical advances~\cite{lw}. To perform the lattice simulations,
periodic boundary conditions will be used, so we shall first discuss
how this changes the expectations for the static force  presented
above for the infinite-volume case.

To fix the notation, let us consider the three-dimensional volume to
be a Euclidean hypertorus (or box with periodic boundary conditions
in all directions),  with a coordinate $r$ in the spatial direction
along the brane,  $t$ in the temporal direction along the brane,  and
$z$ along the ``bulk'', perpendicular to the brane. The extents of
each direction (if finite) are denoted by $L_r, L_t, L_z$, 
respectively.  The brane is located at $z=0$. 

The static force is defined in the usual way.  We introduce a rectangular 
Wilson loop $W(R,T;z)$ in the ($r,t$)-plane, at some fixed $z$, of 
size $R \times T$. The force $F(R;z)$ is defined through the potential 
$V(R;z)$ as
\be
 F(R;z) = -\frac{{\rm d}\, V(R;z)}{{\rm d}R} = \lim_{T\to \infty}
 \frac{1}{T} \frac{{\rm d}}{{\rm d}\, R} \ln W(R,T;z) 
 \;, \la{FWilson}
\ee
where we assumed that $L_t = \infty$. 

If we imagine for a moment that the system is homogeneous
and perturbative, a leading order computation in $g^2$ gives
\be
 F(R;z) = g^2 C_F \frac{1}{L_z} \sum_{p_z}
 \frac{\sinh[p_z(L_r/2-R)]}{2 \sinh(p_z L_r/2)}
 \;, \la{FRz}
\ee
where $C_F$ is the quadratic Casimir of the fundamental
representation, $C_F \equiv (N_c^2-1)/2N_c$,  and $p_z= 2 \pi n/L_z$,
with $n$ an integer.   This result is exact for an Abelian theory
(Coulomb law), but  in a non-Abelian confining theory it is only
valid at short distances and, in general, the periodicity
it displays will be
lost at large distances. 
Nevertheless, as we will see,  various interesting limits can
be obtained from this simple expression. 
In the following we set $N_c = 2$ as in the actual  simulations, so
that $C_F = 3/4$.

Let us consider three limiting cases:

\paragraph{Infinitely thin brane: the symmetric phase.}
In the confining symmetric phase, the force is analytically computable
only at small distances (the Coulomb part). 
Because of the mass gap, we can set 
$L_z \to \infty$ in~\eq\nr{FRz}, so that the sum $\sum_{p_z}$
becomes an integral. Then,
\ba
 F(R;z) & = &
 \fr34 \frac{g^2}{2\pi R} ~~~~~~~,~~~~~~~~~~
 R \ll (g^2 C_A)^{-1} \;, \\
 & = & c_1 (g^2 C_A)^2   ~~, ~~~~~~~~~~
 (g^2 C_A)^{-1} \ll R \ll L_r/2\;, \la{3dconf}
\ea
where $\sigma = c_1 (g^2 C_A)^2$ is the string tension
of 3d SU(2) Yang-Mills theory with adjoint matter, and 
$C_A = N_c$ is the quadratic Casimir of the adjoint
representation. Without
adjoint matter, the constant $c_1$ is 
determined numerically to be $\approx 0.0281(3)$ in
the continuum limit for $N_c = 2$~\cite{mt}, and the inclusion of adjoint
matter leaves the value practically unchanged~\cite{hp}. 

\paragraph{Infinitely thick brane: the broken symmetry phase.} 
In the opposite limit of a 3d broken symmetry 
phase, the three isospin components
of the vector bosons 
convert into two massive vector bosons, of mass 
$m_W = g \langle \chi^3 \rangle$, and one massless vector boson, the photon. 
In 3d, however, the photon becomes  massive~\cite{p2} 
via interactions with monopoles~\cite{th,p1},  
\be
 m_\gamma \sim g^{-3/2} m_W^{7/4} \exp\Bigl(-\frac{2\pi m_W}{g^2}\Bigr) 
 \;.
 \la{mgamma} 
\ee
Correspondingly, the potential is computable in weak coupling only up to 
distances $\sim m_\gamma^{-1}$. Assuming again, for simplicity, that
$L_z \to \infty$, we obtain
\ba
 F(R;z) & = &  
 \fr34 \frac{g^2}{2\pi R} ~~ , ~~~~
 R \ll m_W^{-1} \;, \\
 & = &  \fr14 \frac{g^2}{2\pi R} ~~ , ~~~~
 m_W^{-1} \ll R \ll m_\gamma^{-1} \;, \la{powerlaw} \\
 & = & c_2 g^2 m_\gamma ~~ , ~~~
 m_\gamma^{-1} \ll R \ll {L_r}/{2} \;,  
\ea
where $c_2$ is a constant. If $L_r/2 \ll m_\gamma^{-1}$, as is realistically
the case, then at large distances we rather encounter the fully perturbative 
behaviour for the 
photon following from~\eq\nr{FRz} (still in the limit $L_z \to \infty$), 
\be
 F(R;z) = \fr14  \frac{g^2}{2 L_r \tan(\pi R/L_r)} 
 ~~ , ~~~ m_W^{-1} \ll R
 \;. \la{3dcoulomb}
\ee

\paragraph{Brane of a finite width.} 
Consider finally the case of our actual interest, a domain wall of some
finite effective width, $\sim L_\rmi{brane}$. The behaviour should now 
interpolate between the two limits encountered above. The 
input parameter determining the width is $\alpha = \ell g^2$ as 
defined in~\eq\nr{alpha}. Independent of $\alpha$, at very 
small distances we still have
\be
  F(R;z) = \fr34 \frac{g^2}{2\pi R} \; , ~~~ R \ll m_W^{-1} \;. 
\ee
What happens at large distances, on the other hand, 
depends on whether there is a zero-mode 
or not. If the zero-mode exists, we may expect that its contribution 
is according to \eq\nr{FRz}, where $L_z$ is now finite and replaced with 
$L_\rmi{brane}$, and we take $p_z = 0$, and only one of the isospin
components contributes so that $C_F  = 3/4 \to 1/4$, 
\be
 F^{\rmi{(0)}}(R;|z|\lsim L_\rmi{brane}) 
 \sim \frac{g^2}{4} \frac{1}{2 L_\rmi{brane}}
 \biggl(1 - \frac{2 R}{ L_r}\biggr) 
 \;. \la{zero}
\ee
This would be the behaviour if the confining phase outside the brane 
acted effectively as a Neumann (derivative of field vanishes), 
rather than Dirichlet (field itself vanishes), boundary condition.  
The first massive mode, on the other hand, contributes as 
\ba
 F^{\rmi{(1)}}(R;|z|\lsim L_\rmi{brane})  
 & \sim & \frac{g^2}{4} \frac{1}{2 L_\rmi{brane}}
 \frac{\sinh[m_1(L_r/2 - R)]}{\sinh(m_1 L_r/2)}
 \\ 
 & \approx & \frac{g^2}{4} \frac{\exp(-m_1 R)}{2 L_\rmi{brane}} \;, 
 ~~~ R \ll L_r/2\;, \quad m_1 L_r \gg 1
 \;, \la{m1}
\ea
where the mass $m_1$ is assumed non-zero. This would also 
be the full
behaviour with Dirichlet boundary conditions, 
in which case a zero-mode is excluded. 

Let us emphasize that the linear in $R$ behaviour of the force in
\eq\nr{zero} is a characteristic of the Abelian theory.  If the
2d low-energy effective theory were non-Abelian, 
the long-distance force would still be constant as
in \eq\nr{3dconf}, with  exponentially
small finite-volume corrections.
This difference is caused by the finite periodic extent 
of the box in the $R$-direction --- in infinite volume both
the 2d Abelian and non-Abelian gauge theories display linear
confinement (constant force) \`a la \eq\nr{3dconf}.

\section{Lattice formulation}
\la{se:latt}

\subsection{Discretised action}

In order to test the behaviour at large distances, so as for instance to 
distinguish between \eqs\nr{zero} and \nr{m1}, we study the system
on the lattice. The discretised Lagrangian corresponding to \eq\nr{leff} is
\ba
 {\mathcal L}_{\rm latt}
 &=&
 \frac{1}{a^4g^2}\sum_{k,l}\tr\left[{\bf 1}-P_{kl}(x)\right]\nn
 &&+
 \frac{2}{a^2}\sum_k
 \left[\tr \chi^2(x)-\tr \chi(x)U_k(x) 
 \chi(x+a\hat e_k)U_k^\dagger(x)\right] \nn
 &&+
 m^2_\rmi{bare}(z)\tr[\chi^2]+\lambda(\tr[\chi^2])^2
 \;, \la{lattL}
\ea
where $a$ is the lattice spacing, $U_k(x)=\exp[iagA_k(x)]$,
$\hat e_k$ is a unit vector in direction $k$, 
and $P_{kl}$ is the plaquette:
\be
 P_{kl}(x)=U_k(x)U_l(x+a\hat e_k)
 U_k^\dagger(x+a\hat e_l)U_l^\dagger(x)
 \;.
\ee
The extents of the box are denoted by  
$L_r = a N_r, L_t = a N_t, L_z = a N_z$, and the lattice volume
by $V = N_r N_t N_z$.
The lattice spacing is
expressed through the dimensionless combination 
\be
 \beta \equiv \frac{2 N_c}{a g^2}
 \;.
\ee

In order for physics to be the same as with the continuum 
Lagrangian in~\eq\nr{leff}, 
the lattice theory needs to be renormalised. In three dimensions, 
the only parameter including divergences is $m^2$, and the 
divergences can be computed exactly, close to the continuum 
limit~\cite{framework,contlatt}: 
\ba
 m_\rmi{bare}^2(z) &\equiv& m^2(z) + \delta m^2 (\bmu), \\
 \delta m^2 (\bmu) 
 &=& - \Bigl[2g^2C_A + \lambda(d_A+2)\Bigr] \fr\Sigma{4\pi a} \nn
 & & +\fr1{16\pi^2} \lk  2\lambda(d_A+2)\(\lambda-g^2C_A\)
 \(\ln\fr6{a\bmu} +\zeta\) -2g^2C_A\lambda(d_A+2)
 \(\fr{\Sigma^2}4-\delta\) \rd \nn 
 & & \ld -g^4 C_A^2 \( 
 \frac{5}{8}\Sigma^2+\(\frac{1}{2}-\frac{4}{3C_A^2}\)\pi\Sigma
 -4(\delta+\rho)+2\kappa_1-\kappa_4
 \) \rk \;, \la{dmassL}
\ea
where the constants
$\zeta$, $\delta$, $\rho$, $\kappa_1$, $\kappa_4$ and $\Sigma$
have been defined in Refs.~\cite{framework,contlatt}, 
$C_A = N_c$, $d_A = N_c^2 - 1$, 
and $\bmu$ is the scale
parameter of the $\msbar$ scheme. The couplings
$g^2, \lambda$ require no renormalisation, but the approach 
to the continuum limit could be improved by computing 
corrections of order ${\cal O}(a)$~\cite{moore_a}.
As already mentioned, we choose $\bmu = g^2$ in the following. 
Various condensates, such as $\tr [\chi^2]$, 
also require additive renormalisation
(multiplicative wave function renormalisation effects
are ${\cal O}(a)$~\cite{moore_a}):
\ba
 \langle \tr [\chi^2] \rangle_\rmi{$\msbar$} & = &
 \langle \tr [\chi^2] \rangle_\rmi{bare} - d_A \frac{\Sigma}{8\pi a} 
 - d_A C_A \frac{g^2}{16\pi^2}
 \biggl( \ln\fr6{a\bmu} + \zeta + \frac{\Sigma^2}{4} - \delta
 \biggr) 
 \;. 
\ea

\subsection{Discretised observables}

The basic object we employ on the lattice is a Polyakov loop, 
\be
 P(R;z) \equiv \tr \Bigl[
 \Pi_{n_t = 0}^{N_t - 1} U_t(R,n_ta,z) \Bigr]
 \;,
\ee
and the correlation function of two Polyakov loops is
\be
 C(R;z) \equiv \sum_{n_r = 0}^{N_r - 1}
 \Bigl\langle
 P^\dagger (n_r a;z) P(n_r a + R;z)
 \Bigr\rangle
 \;. 
 \la{Crt}
\ee
The general structure of this correlator is
(\cite{lw} and references therein) 
\be
 C(R;z) = w_1 e^{-L_t V(R;z)} + w_2 e^{-L_t [V(R;z) + \Delta E]} + ...
 \;,  \la{FCfull}
\ee
where $w_i$ are numerical coefficients and 
$\Delta E$ is a mass gap. Therefore the static force, 
to be denoted by $F_P(R;z)$ when extracted from the Polyakov loop
correlator, can be defined as
\be
 F_P(R;z) \equiv \frac{1}{L_t} \frac{{\rm d}}{{\rm d}R} \ln C(R;z) 
 \;, 
 \la{FCcont}
\ee
and the difference between $F_P(R;z)$ and $F(R;z)$ is
exponentially small if $\Delta E > 0$ and $L_t$ is large enough. 
On the lattice we use a discretised version of~\eq\nr{FCcont}, 
\be
 F_P(R+\frac{a}{2};z) \equiv
 \frac{2}{aL_t} \frac{C(R+a;z)-C(R;z)}{C(R+a;z)+C(R;z)}
 \;.
 \la{FClatt}
\ee

The reason for using Polyakov loop correlators instead of the more
common Wilson loop to obtain the static force, 
is simply that a sufficient numerical accuracy is
easier to reach, thanks to the advanced numerical techniques developed by
L\"uscher and Weisz~\cite{lw}
(and optimised for the present system by us).  The
measurement method will be described below.  This
advantage does not come without a price, however: if the system is not
confining, then the error made by using \eq\nr{FCcont} rather than the full
expression in~\eq\nr{FCfull} may in general not be small, and the
correct result is only obtained after an extrapolation to the limit of
a large $L_t$.  Nevertheless, even data obtained at finite $L_t$ will
correctly show whether the behaviour of the force is constant,\footnote{%
 At large distances the behaviour of $F_P$ deviates from a constant, 
 due to the periodic boundary conditions. Based on the effective string
 picture, the behaviour expected is the same as in 2d~\cite{Rusakov}, 
 $F_P(R;z) = \sigma \tanh[\sigma L_t(\fr{L_r}{2} - R)]$ for large $L_r$,  
 which approaches a step function for $L_t\rightarrow \infty$.
 In practice, however, the error bars are reasonably 
 small only for $R \ll L_r/2$, 
 where $F_P$ can be well approximated with a constant. 
 } 
linear, or exponential in $R$, thus allowing to 
distinguish between \eqs\nr{3dconf}, \nr{zero}, \nr{m1}. 

In the following, we will always
refer to the object defined by \eqs\nr{FCcont}, \nr{FClatt} as the
static force, while keeping the reservations just spelled out in mind,
and therefore checking in the end explicitly for the stability of our
results with respect to variations of $L_t$.

\subsection{Parameters and systematics}

The model of~\eq\nr{lattL} has previously been studied with 
lattice simulations in Refs.~\cite{hpst,adjoint}, for the case of a 
homogeneous mass parameter, corresponding to $\alpha = 0$
or $\alpha = \infty$. The system was observed to have a non-trivial
phase diagram in the space of the continuum parameters $x,y$, 
defined in~\eq\nr{alpha}: for $x < x_c \approx 0.3$, there is 
a first order phase transition separating the broken symmetry 
phase at $y < y_c$ from the symmetric phase at $y > y_c$. For 
$x > x_c$, on the other hand, there is no phase transition, meaning that 
the two phases are analytically connected. This is possible 
since there are monopoles in the broken symmetry phase~\cite{th,p1}
which replace a possible order parameter, the mass of a photon related
to an unbroken U(1) symmetry, by a small but non-vanishing mass
(given in~\eq\nr{mgamma}) 
related to a pseudoscalar particle~\cite{p2}. 
The situation would be 
different with the gauge group SU(3), for instance, in which case
there is an additional discrete global symmetry $\chi \to -\chi$
which gets broken at $y = y_c$; thus there always is a genuine phase
transition of some kind~\cite{su3}.

In terms of the parameters of~\eq\nr{alpha}, we choose here to 
study the value $x=0.20$, which implies a first order transition at 
$y_c \approx 0.147$~\cite{adjoint} (provided one is close to the continuum
limit, i.e., $\beta \gg 1$). The motivation is to have a clear 
distinction between the two phases, 
and to avoid complications  owing to the presence of monopoles, 
which are more important in the crossover region.
The parameters $y_1,y_2$ of \eq\nr{alpha} are chosen on the two 
sides of the transition, 
\be
 y_1 = -0.667
 \;, \quad
 y_2 = 2.0
 \;. 
\ee
The width of the domain wall is varied
in small intervals in the range $\alpha = \ell g^2 = 0.0...5.3$, 
and we also have data for $\alpha = \infty$.

In order to control finite lattice spacing and finite volume artifacts, 
we have carried out simulations at two values of 
$\beta$, $\beta = 6,9$, 
and with a series of volumes in the range 
$V = N_r N_t N_z = 24^3$...$48^3$. Experiences from previous 
studies of the same system~\cite{hpst,adjoint,hp} as well as from 
glueball computations with pure SU(2) gauge theory
in three dimensions~\cite{mt} suggest that such values
are already safely in the scaling region.

The update algorithm used for the simulations is a combination
of one heat bath update cycle followed by four overrelaxation updates
for both the gauge and the adjoint scalar fields.  
More details concerning the implementation 
can be found in Ref.~\cite{adjoint}.

\subsection{Measuring the Polyakov loop correlation functions}

In a confining system at large $L_t$ the Polyakov loop correlation
function is a very ``noisy'' observable: the magnitude of the
correlation function is $\sim \exp(-\sigma L_t R)$, whereas the noise
is always of order unity.  Thus, without any advanced techniques huge
statistics is needed.  We employ here a modification of the multilevel
approach presented in Ref.~\cite{lw}.

To summarize our method, let us 
again consider a lattice of size 
$N_r\times N_t\times N_z$, 
where the Polyakov loops are oriented along the $t$-direction and the
correlation function is to be measured in the $r$-direction.  One measurement
cycle works as follows:
\begin{itemize}
\item[1.] Divide the
  lattice into $N_t$ sublattices by {\em freezing} the following variables:

  (a) gauge link variables $U_r$ and $U_z$ and adjoint scalars $\chi$ 
   which reside on the $(r,z)$-planes located
   at $t/a=0,2,\ldots N_t-2$,\footnote{%
     We assume that indices start from zero, 
     i.e., $t/a=0\ldots N_t-1$, for example.}
   and

  (b) $U_t$, $U_z$ and $\chi$ on $(t,z)$-planes at $r=0$ 
  and $r/a=N_r/2$.
  
\item[2.] Perform $N$ update sweeps on the non-frozen variables within
  all the sublattices.  Because the boundaries of the sublattices are
  fixed, we can generate a valid full lattice configuration by
  choosing any of the $N$ configurations from each sublattice, giving
  a total of $N^{N_t}$ lattice configurations.
  
\item[3.] The Polyakov loop correlation function is measured so that
  one of the loops is located in the interval $0 < r_1/a < N_r/2$ and
  the other at $N_r/2 < r_2/a < N_r$.  This gives effectively $N^{N_t}$
  measurements; the exponential growth with $N_t$ compensates for the
  exponential decrease in the signal. 

\item[4.] In practice, one improved correlation function
  measurement is (we suppress the $z$-coordinate)
  \be
       \bar C(r_2 - r_1) = \bar{P}^\dagger(r_1) \bar{P}(r_2)
  ~,~~~~
   ~~ 
    \bar{P}(r) = \tr\left[\bar{S}(0,r)
      \bar{S}(2a,r)\ldots\bar{S}((N_t-2)a,r)\right] \;,
  \ee
  where $\bar S$ is the average of the 2-link piece of the
  Polyakov loop within one sublattice:
  \be
      \bar{S}(t,r) = \fr{1}{N} \sum_{i=1}^N U_t^{(i)}(t,r)U_t^{(i)}(t+a,r).
  \ee
  Clearly, $\bar S$ is an average over $N$, $\bar P$ over
  $N^{N_t/2}$ and $\bar C(r_2-r_1)$ over $N^{N_t}$ measurements.

  The results are further averaged over $r_1$ while 
  $R=r_2-r_1$ is kept fixed.   However, $r_1$ and $r_2$ should not
  be too close to the fixed $r=0$ and
  $r/a=N_r/2$ planes, which would diminish the configuration-by-configuration
  variation and hence the noise reduction.  Thus, the technique works best
  when $R/a$ is largest, $\sim N_r/2$, just where it is needed.  

\end{itemize}
The parameter $N$ is to be optimized for  the physical situation in
question; the range we  used is $N\sim 100-300$.   Naturally, the $N$
configurations are far from independent; thus, standard update cycles
must be performed too.  Nevertheless, the method achieves the goal of
strongly reducing the statistical noise. Clearly, maximizing the
number of sublattices achieves the highest averaging.  However,
$t$-slices of thickness 1 (which would yield $N^{2N_t}$ measurements)
are not very useful, because the variables within blocks would be too
strongly locked in place. 

The difference between the method  above and the one presented in
Ref.~\cite{lw} is that the latter does not implement the freezing of
the planes in step 1(b) above.  The number of measurements is thus
cut  to $N^{N_t/2}$; on the other hand a larger optimal $N$ tends to 
compensate for this.  In our case the existence of the scalar field
$\chi$ in general lowers the useful values for $N$ --- indeed, in the
broken phase the whole improvement becomes largely unnecessary.

\vspace*{0.5cm}

Let us end this section with another technical comment. Before
starting the  measurements, the system needs to be ``thermalised''.
That is, one starts from some initial configuration (for instance
``cold'', whereby all fields are frozen to values corresponding to
the classical minimum of the action, or ``hot'', whereby all fields
have random values) and carries out updates until the system has
reached a typical ``fluctuating'' configuration, which should be
independent of the initial one. It now turns out that with a ``hot''
start, the  thermalisation process is anomalously slow. The reason is
that the system can  contain U(1) vortices and anti-vortices 
penetrating through the domain wall. Since there is no net magnetic
flux, all the vortices and anti-vortices must eventually evaporate,
but microscopic updates are very slow in  achieving such a global
change in practice. With a ``cold'' start, on the contrary,  there is
no problem with thermalisation, and it is thus the method of choice
for this system.

\section{Results}
\la{se:results}


\begin{figure}[tb]

\centerline{
    \psfig{file=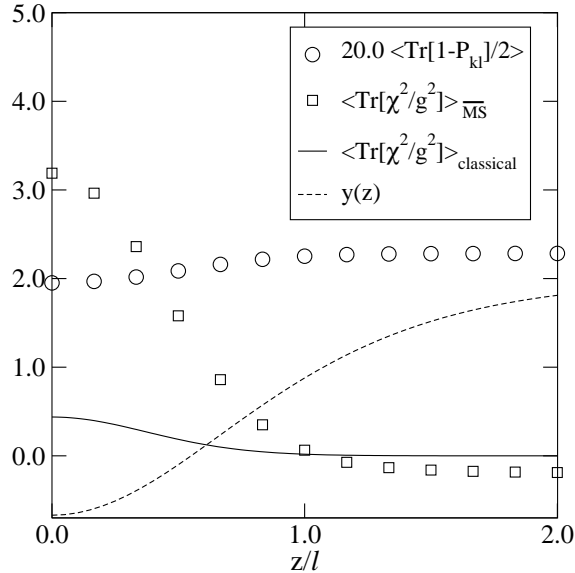,angle=0, width=7.5cm}
}


\caption[a]{Various condensates, 
as well as the mass parameter $y(z)$,  as a function of $z/\ell$ 
at $\alpha = \ell g^2 = 2.667$, $\beta=9$, $V =  48^2 \times 32$. 
In a homogeneous
system, the phase transition point is at
$y(z) = y_c \approx 0.15$~\cite{adjoint}.
The classical prediction for the scalar condensate, obtained
by solving the equations of motion numerically, is also shown. 
}

\la{fig:conden}
\end{figure}

We now move on to discuss our numerical results. 
To start with, we show in \fig\ref{fig:conden} the actual 
structure of the domain wall for typical parameter values, 
in terms of local condensates. It is seen that, judging 
by the eye, the width of the domain wall is indeed well determined
by the input parameter $\ell$, as defined through~\eq\nr{m2z}. 
There is quite a difference, though, in the magnitudes of the lattice
value of the condensate (square boxes) and the tree-level value
(continuous line). The difference can be explained by the fact that with 
our choice of parameters higher order corrections to condensates are
of the same order as the tree-level value, so that perturbation theory cannot 
be trusted. We have checked (in the limit of a thick domain wall) that 
an account of 1-loop effects makes the difference considerably smaller.

\begin{figure}[tb]

\centerline{
  \psfig{file=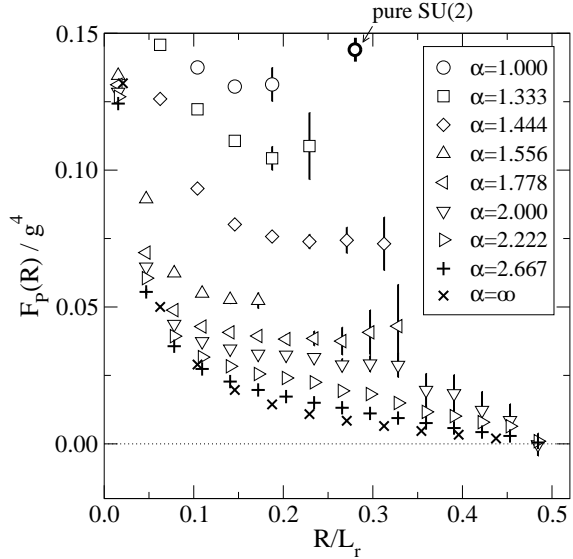,angle=0, width=7.5cm}%
}


\caption[a]{The static force for various values of $\alpha$, 
at $\beta = 6$, as a function of $R/L_r$.
When not visible, the error bars are smaller than the symbol sizes. 
The volume is 
$V = 24^3$ for $\alpha = \infty$, and $V=32^2 \times 24$ otherwise. 
The pure gauge value at $\beta = 6$ is taken from Ref.~\cite{mt}.} 

\la{fig:hom}
\la{fig:force_ff}
\end{figure}

Our basic observable, the force $F_P(R)$ in the central plane $z=0$, 
is shown in~\fig\ref{fig:force_ff}, 
for various values of $\alpha = \ell g^2$.
For $\alpha = 0$, corresponding to 3d confinining behaviour, the force 
becomes constant (or rather tanh-like) at intermediate distances, 
representing the 3d string tension.  
Measurements are relatively difficult, since the Polyakov loop correlation
function decays rapidly with $R$. As $\alpha$ is increased,
the interior of the domain wall starts to look more and more like
a homogeneous broken symmetry phase,  and 
the plateau moves down, eventually 
disappearing completely. The force then resembles 
the behaviour observed in the 3d broken symmetry phase, 
\eq\nr{3dcoulomb}, corresponding to $\alpha = \infty$.

\begin{figure}[tb]

\centerline{
  \psfig{file=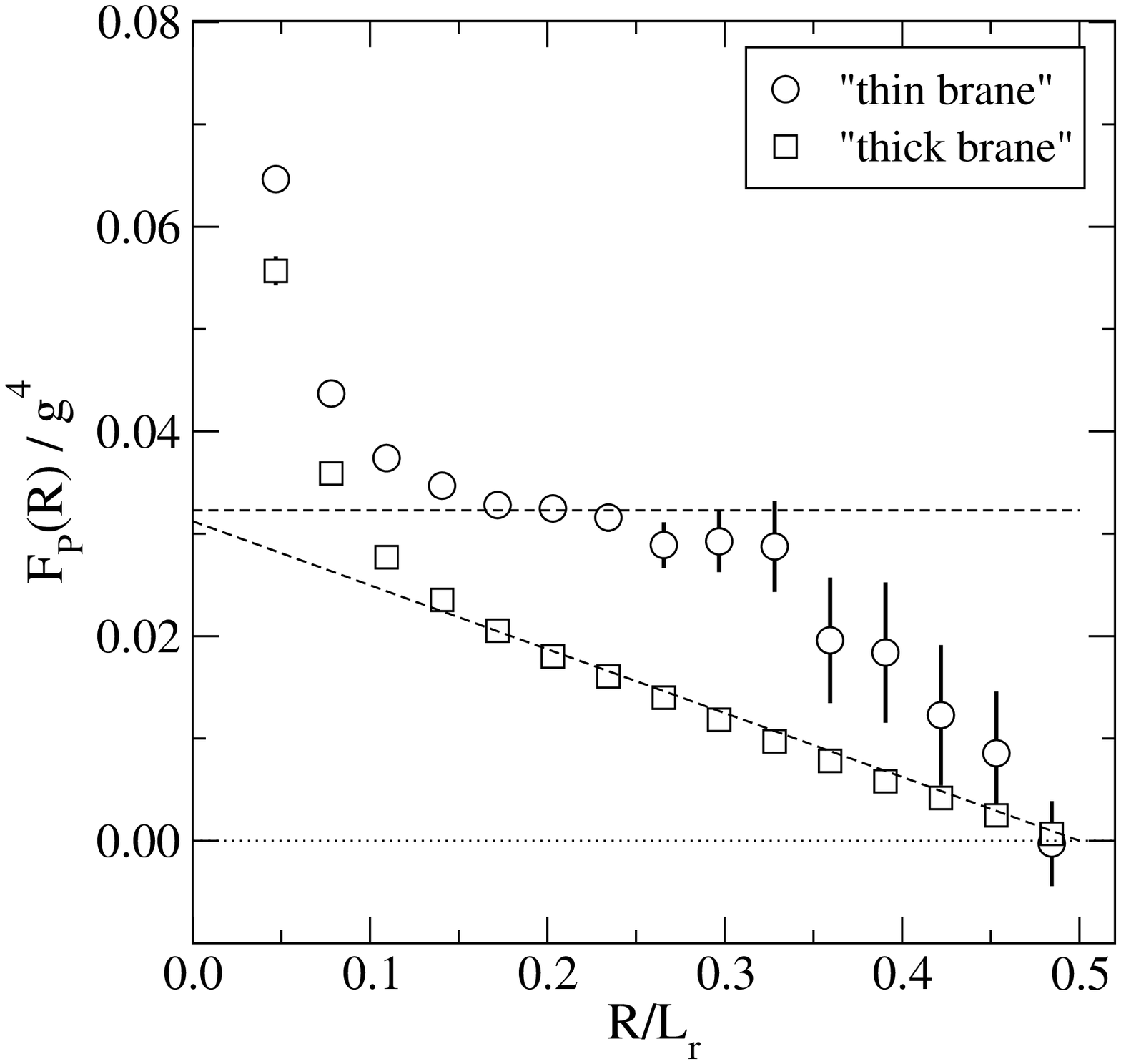,angle=0, width=7cm}%
  ~~\psfig{file=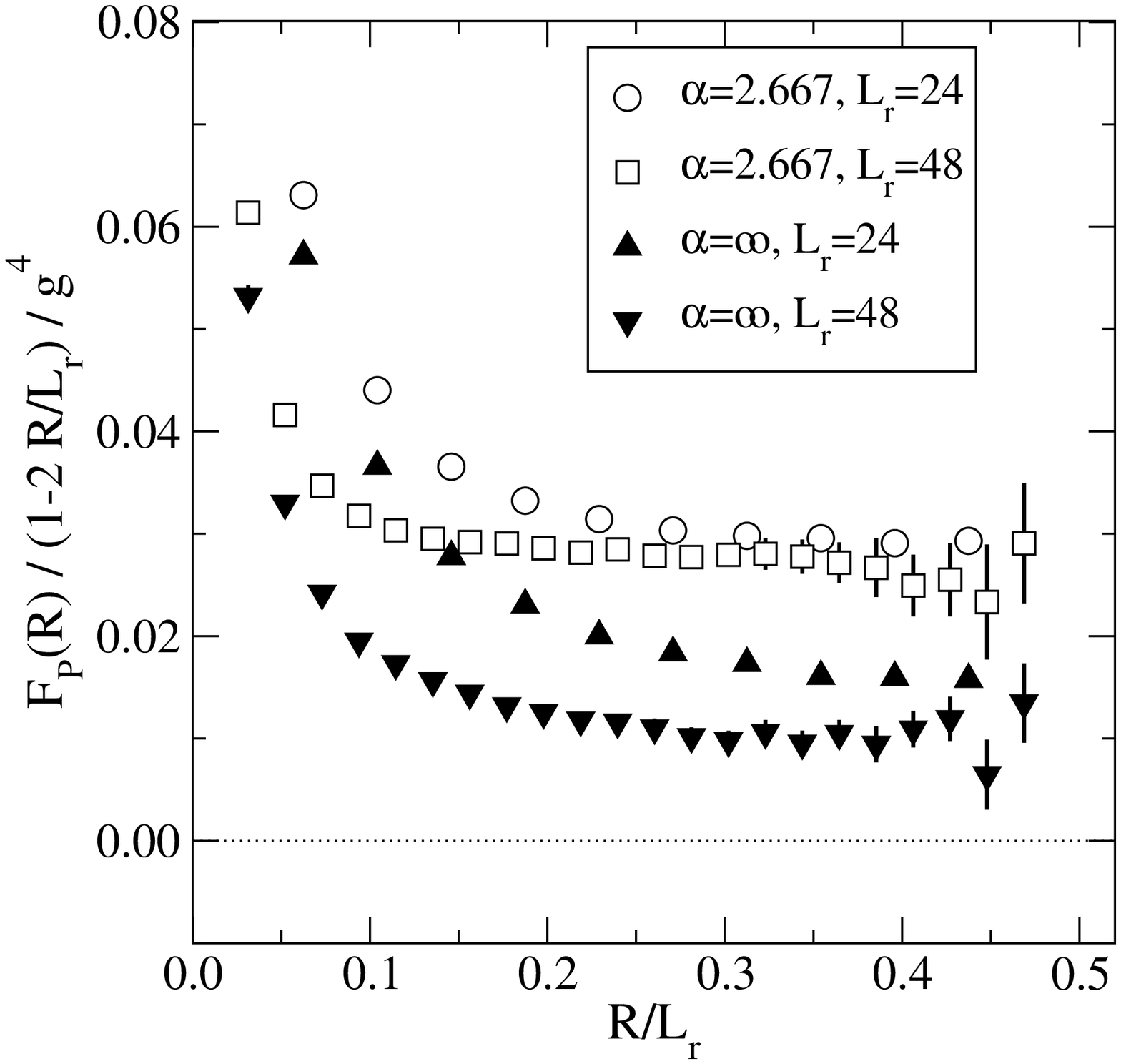,angle=0, width=7cm}%
}


\caption[a]{Left:
The static force in the central plane in two different runs,
at  $\ell g^2=2$, $\beta=6$, $V=32^2\times 24$. The system 
can be in one of two metastable states, resulting in different
qualitative behaviours for the force. These are referred to as 
the ``thin brane'' and the ``thick brane'', because the former
has the functional form appearing in the limit $\ell \to 0$, 
the latter in the limit $\ell \to \infty$.
Right: The ``thick brane''
static force, divided by $(1-2R/L_r)$, at 
different values of $L_r$, compared with results for 
the homogeneous broken symmetry phase. The parameters 
are $\beta$ = 6,  $V$ = $L_r \times 24^2$.}

\la{fig:ell3}
\la{fig:force6v}
\end{figure}

The quantity we would like to extract from curves of the
type in \fig\ref{fig:force_ff} will be referred to as the
``string tension''. There are, however, two qualitatively different 
regimes for the behaviour observed: for small $\alpha$, the domain wall is 
essentially in the symmetric phase and we extract
the string tension from the plateau in $F_P(R)$ at $R \ll L_r$.
On the other hand, for larger $\alpha$
the domain wall is in the broken symmetry phase, and the string tension
is extracted from the coefficient 
of a ``linear term'' in $F_P(R)$, that is, from $F_P(R)/[1- 2 R/L_r]$ at 
$R \approx L_r / 2$. 
It turns out that, as a remnant of the first order transition 
experienced by the homogeneous system at $x = 0.20$,  
for certain values of $\alpha$ the central plane can even be in one 
of two metastable branches, exhibiting these two patterns.
This situation is illustrated in~\fig\ref{fig:force6v}(left).

While a force constant in $R$ at intermediate distances
is a signal of non-Abelian 3d confinement, 
a behaviour linear in $1-2 R/L_r$ as $R\rightarrow L_r/2$  
still allows for two 
different interpretations: 3d Coulomb phase, characterised
by~\eq\nr{3dcoulomb}, and 2d Coulomb phase, characterised by~\eq\nr{zero}. 
One can differentiate between the two by
approaching $R \approx L_r / 2$, where
\ba
 \frac{\sigma}{g^4} \equiv \lim_{R\to \frac{L_r}{2}}
 \frac{1}{g^4} \biggl[ \frac{F_P(R)}{1-2R/L_r} \biggr] & = & 
 \frac{3\pi}{16 g^2 L_r} \;, ~~~~~~ \mbox{3d U(1)} \;, \la{probe1} \\
 & = & \frac{1}{8 g^2 L_\rmi{brane}} \;, ~~~ \mbox{2d U(1)}
 \;. \la{probe}
\ea
Therefore, if we increase the system size $L_r$, the existence
of a 2d zero-mode is signalled by the force approaching 
a constant,\footnote{%
  Let us stress again that 
  the existence of a string tension in an Abelian theory
  is specific to 2d.
  }
while for a 3d Coulomb phase the apparent ``string tension'' vanishes
in the infinite volume limit as $\sim 1/L_r$, and the potential becomes 
a power law, \eq\nr{powerlaw}. These different
qualitative behaviours are illustrated in~\fig\ref{fig:force6v}(right).
Thus, whether dimensional reduction takes place or not
can be seen by monitoring the dependence of the observable
in~\eq\nr{probe1} on the box size $L_r$. 

\begin{figure}[tb]

\centerline{
     \psfig{file=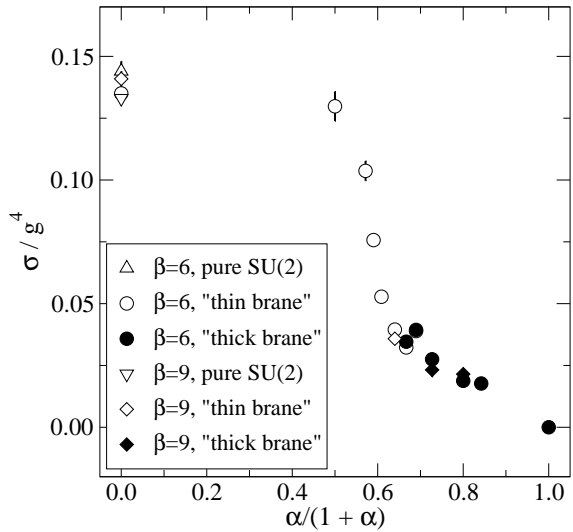,angle=0,width=7.5cm}
}


\caption[a]{The string tension, $\sigma/g^4$, 
as a function of $\alpha/(1+\alpha)$.
Pure gauge values are from Ref.~\cite{mt}.
The physical volume 
for the three smallest $\alpha$'s is $g^6 a^3 V = 16^2 \times 10.67$,
otherwise $g^6 a^3 V=21.33^2 \times 16$.
The labels ``thin brane'' and ``thick brane'' refer to whether 
the functional form of the data points resembles
the upper or lower curve in~\fig\ref{fig:ell3}(left), 
which in turn determines the fit ansatz employed (see the text).}

\la{fig:sig_ell}
\la{tab:sig_ell}
\end{figure}

Results for the string tension, extracted either from a plateau
in the static force $F_P(R)$, 
or as defined in~\eq\nr{probe1}, are shown as a function 
of $\alpha$ in~\fig\ref{fig:sig_ell}. We observe that the behaviour
interpolates between the 3d confinement and 3d Coulomb behaviours. For 
narrow domain walls, or small $\alpha \lsim 2.0$, the system behaves
as a 3d confinement phase (``thin brane''). 
For larger $\alpha$, it is dominated by
a 2d zero-mode (``thick brane''), 
with a string tension scaling as $\sim 1/L_\rmi{brane}$, 
as in~\eq\nr{probe}. Note that
the system goes over into the ``thick brane'' regime only when
the string tension is a factor 3...4 smaller than in the bulk.
Increasing $\alpha$ further, 
the string tension continues to decrease, but the system 
starts simultaneously to look more and more like a 3d Coulomb phase, 
to which it finally goes over as $\alpha\to\infty$.

\begin{table}

\vspace{0.5cm}

\centerline{
\begin{tabular}{cccc}
\hline
 \multicolumn{2}{l}{$\alpha = 2.667$} & 
 \multicolumn{2}{l}{~~~~~$\alpha = \infty$}
\\ \hline
 $V = L_r \times 24\times 24$  & $ \sigma / g^4$  & 
 ~~~~~$V = L_r \times 24\times 24$  & $ \sigma / g^4$  
\\ \hline
 $ L_r=24$   & 0.029(1) &
 ~~~~~$ L_r=24$   & 0.0160(5) \\ 
 $ L_r=36$   & 0.031(2) &
 ~~~~~$ L_r=36$   & 0.0110(7) \\ 
 $ L_r=48$   & 0.028(1) &
 ~~~~~$ L_r=48$   & 0.0098(5) \\ 
\hline
 $V=32\times L_t\times 16$  &  $ \sigma / g^4$ & & \\
 \hline
 $L_t=16$  &  0.028(2) & &  \\   
 $L_t=24$  &  0.026(2) & &  \\    
 $L_t=32$  &  0.030(3) & &  \\    
 $L_t=40$  &  0.028(2) & &  \\     
\hline
 $V = 32\times 32 \times L_z$  &  $ \sigma / g^4$ & & \\
\hline
 $L_z=16$  &  0.030(3) & &  \\ 
 $L_z=24$  &  0.028(2) & & \\ 
 $L_z=32$  &  0.029(1) & & \\ 
\hline
\end{tabular}}

\caption{Finite-volume scaling of $\sigma/g^4$ as 
defined in \eq\nr{probe1}, 
at $\beta=6$, $\alpha = 2.667$ (left), and $\alpha = \infty$ (right).}

\la{tab:mp_vol}
\end{table}

As discussed above, to make a rigorous distinction between a 2d
Coulomb and 3d Coulomb behaviour, one has to carry out a finite-size
scaling study, as in~\fig\ref{fig:force6v}(right).
The corresponding $\sigma$'s are
shown in the top part of Table~\ref{tab:mp_vol}: it is again seen
clearly that
at $\alpha = 2.667$ the string tension is indeed independent of 
$L_r$, while at $\alpha = \infty$ it is not. 

Table~\ref{tab:mp_vol} also serves to demonstrate that our 
results are independent of the extents of the system in all  
the directions, for a given lattice spacing. 
The independence of the lattice spacing, for a given physical 
volume, is demonstrated by~\fig\ref{fig:sig_ell}.

\begin{figure}[tb]

\centerline{
  \psfig{file=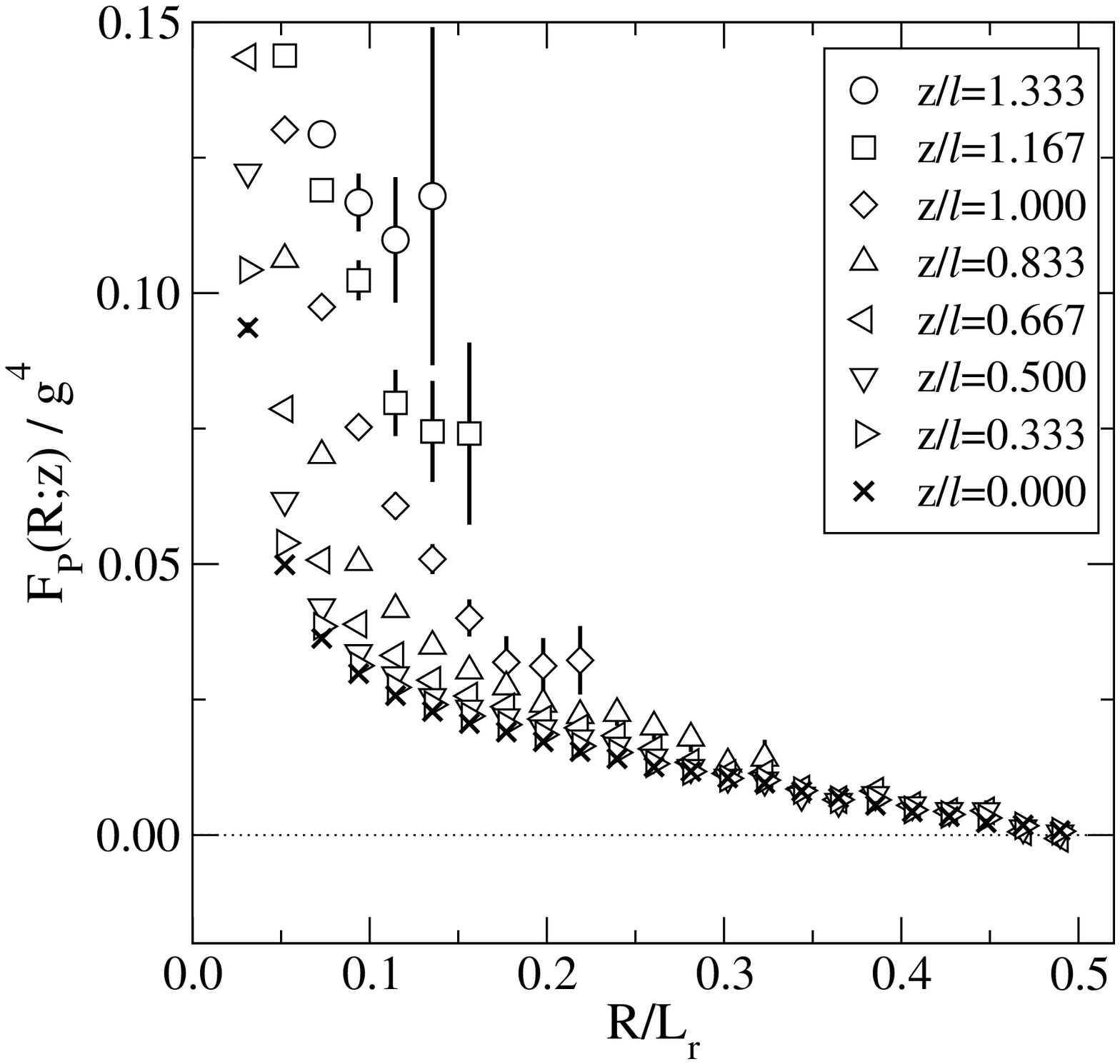,angle=0, height=6.5cm}%
  ~~\psfig{file=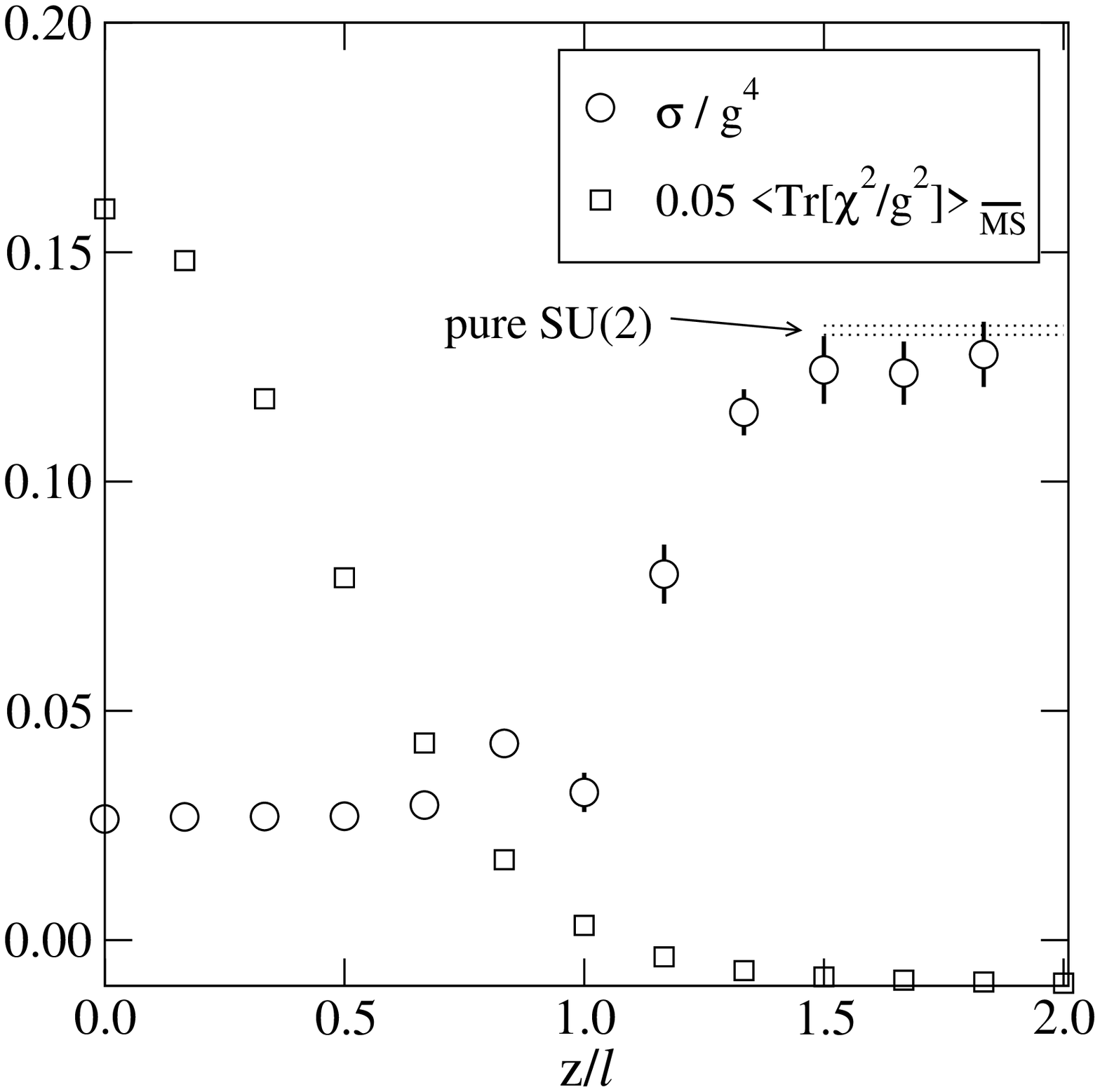,angle=0,height=6.5cm}%
}


\caption[a]{Left: The static force in various planes, for 
$\ell g^2=2.667,~\beta=9,~V=48^2 \times 32$. 
Right: The corresponding string tensions, $\sigma/g^4$, 
together with the scalar condensate. 
For $z/\ell\ge1.0$ ($z/\ell < 1.0$), the fit is of the ``thin brane''
(``thick brane'') type
in \fig\ref{fig:ell3}(left).}

\la{fig:force9p}
\la{fig:wf9}
\end{figure}

The discussion so far has been for the force in the central plane, $z=0$.
In \fig\ref{fig:force9p} we show how the force depends on $z$.
The pattern is as expected: the behaviour characteristic of
the properties of the 2d zero-mode is well localised, and outside
of the domain wall the dynamics is that of the bulk theory.

\begin{figure}[tb]

\centerline{
    \psfig{file=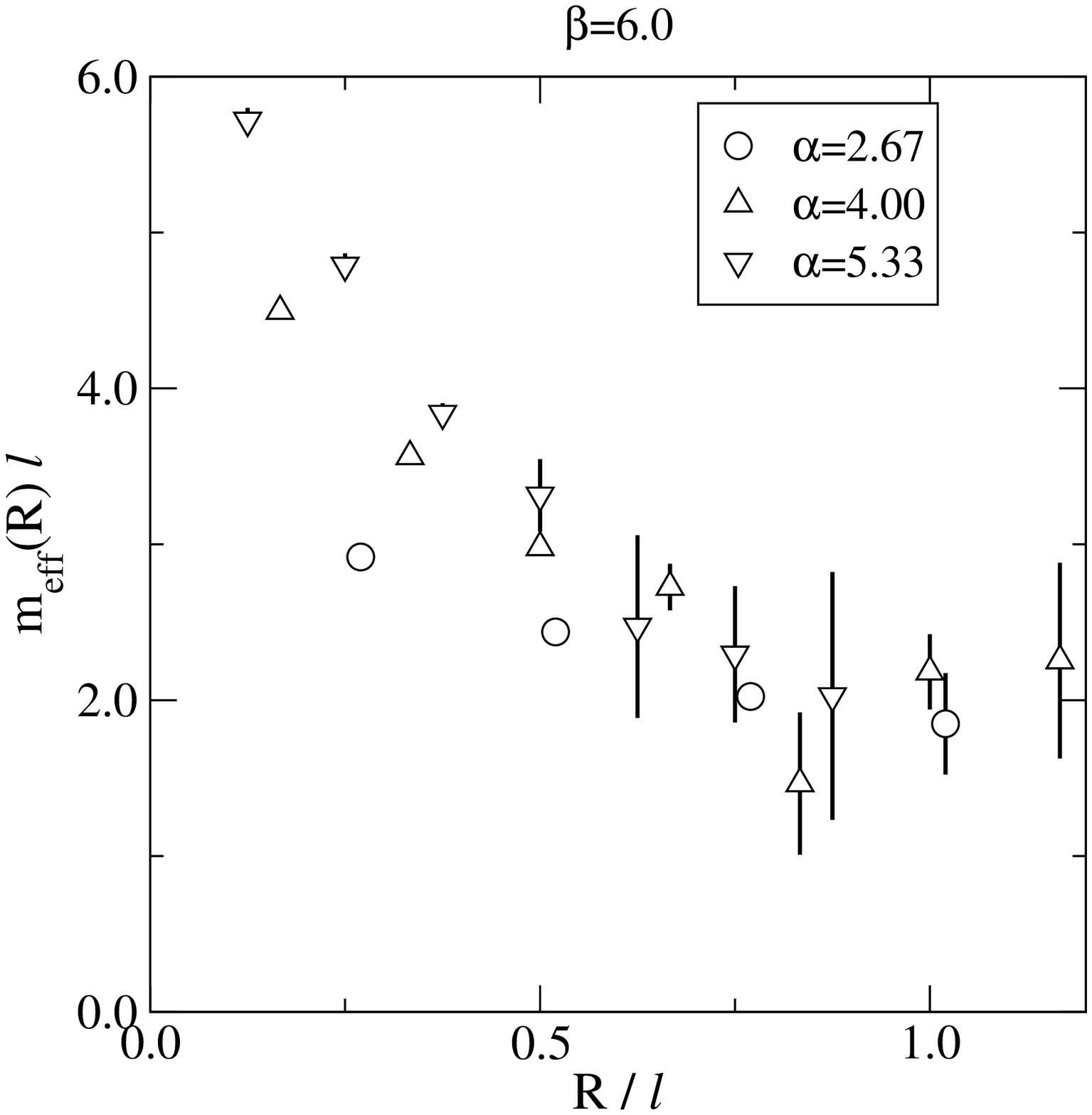,angle=0,width=7.5cm}%
    }


\caption[a]{The observable $m_\rmi{eff}(R)$ 
defined in~\eq\nr{eq:higher}, multiplied by $\ell$, 
as a function of $R/\ell$, for various $\ell$ ($\alpha = \ell g^2$). 
The independence of $\alpha$ at large $R \gsim \ell$ demonstrates that 
$m_\rmi{eff} \sim 1/\ell$.}

\la{fig:higher}
\end{figure}

All the evidence presented so far supports the conjecture that 
the low-energy dynamics on the brane is determined by a localised
2d zero-mode. In order to understand the distance scale at which
corrections to the zero-mode dynamics become important, however, 
one should also determine the mass gap to the higher modes
in the central plane. 
To achieve this, we define the  quantity
\be
 m_\rmi{eff}(R) \equiv - 
 \frac{F_P''(R)} 
 {[F_P(R)/(1-2R/L)]'}
 \;, \la{eq:higher}
\ee
where $[...]' \equiv {\rm d}[...] /{\rm d} R$.
According to~\eqs\nr{zero}, \nr{m1}, the zero-mode does not 
contribute to this observable, and the leading non-trivial contribution 
at large distances is $m_\rmi{eff}(R) \approx m_1$. 

In~\fig\ref{fig:higher} we plot a discretised version of
$m_\rmi{eff}(R)$, in the dimensionless form $\ell\, m_\rmi{eff}(R)$. 
Thus, if $m_1 \sim 1/L_\rmi{brane} \propto 1/\ell$, we should obtain 
a constant value, independent of $\ell$. This indeed is the 
behaviour observed, within statistical errors. Moreover, the distance
where the plateau value is reached is $R/\ell \sim 1$, consistent
with the scenario that the masses of higher modes still 
also scale as $1/\ell$, 
as expected for ``Kaluza--Klein type'' excitations. 

\section{Conclusions and outlook}
\la{se:concl}

The purpose of this paper has been to test with numerical lattice 
Monte Carlo simulations some basic features of a mechanism proposed 
for gauge field localisation on domain wall defects by Dvali and  
Shifman~\cite{ds}. The mechanism relies on non-perturbative dynamics 
in the ``bulk'' outside of the domain wall, and is thus not easily 
``proven'' to work with analytic methods alone.  

Dvali and Shifman considered originally a 3+1 dimensional SU(2) gauge
theory,  arguing that the low-energy dynamics was that of 2+1
dimensional  U(1) gauge theory. For technical reasons, and since the
main physics arguments  remain essentially unchanged, we simplified
in this paper the setting further, and took as the starting point a
2+1 dimensional SU(2) gauge theory,   with adjoint scalar matter. 

The basic pattern we found can be summarised as follows (cf.\
\fig\ref{fig:sig_ell}).  Suppose the 2+1 dimensional theory has a
large confinement scale $\Lambda$, and consider the dynamics on the
brane. If the thickness of the brane is of order unity with respect
to $\Lambda^{-1}$, the dynamics remains the same as in 2+1
dimensions, but the effective string tension decreases rather rapidly
with the thickness (open symbols in \fig\ref{fig:sig_ell}).  As the
brane is made thicker, the dynamics becomes finally 1+1 dimensional
(closed symbols in \fig\ref{fig:sig_ell}), with the ``confinement
scale'' (or string tension, $\sqrt{\sigma}$) taking over (almost)
smoothly from the 2+1 dimensional value. However the transition
between the two behaviours actually seems to  be discontinuous in our
system (cf.\ \fig\ref{fig:ell3}(left)). Increasing the width further,
$\sqrt{\sigma}$ goes down as $\sim 1/\sqrt{L_\rmi{brane}}$, while the
mass gap to the higher modes goes down as $1/L_\rmi{brane}$. An
effective 1+1 dimensional description with an extremely small
$\sqrt{\sigma} \ll \Lambda$ is only reached for $L_\rmi{brane} \gg
\Lambda^{-1}$, and is then valid only at distances $\gg
L_\rmi{brane}$ which implies a kind of a hierarchy problem: there
have to be two {\em different} large scales compared with the
dynamical energy scales of the zero-mode system (such as masses of
bound states,  or $\sqrt{\sigma}$), namely $L_\rmi{brane}^{-1}$ and
$\Lambda$.  Moreover, scales such as $\sqrt{\sigma}$ are  {\em
larger} than the masses of some of the higher modes,  which means
that the contributions from the latter are not suppressed in
generic infrared observables.

How would this pattern change in 3+1 dimensions? If we consider the  
very scenario studied by Dvali and Shifman, then there is in fact a
clear difference. The reason is that the 2+1 dimensional
U(1) theory no longer has a string  tension associated with a
linearly rising potential (ignoring the exponentially small value
produced by the Polyakov mechanism). 
Thus, repeating our measurements, one
would  indeed expect to observe a qualitative change in the  
low-energy dynamics, once the width of the domain wall is   somewhat
wider than the inverse of the   confinement scale outside the brane:
the closed symbols  in~\fig\ref{fig:sig_ell} would all lie on an (almost)
straight line,  at (almost) vanishing $\sigma$, signalling a 
phase transition between the bulk-like ``thin brane'' and 
localised ``thick brane'' regimes.  

On the other hand,   if we rather consider a non-Abelian case, say a
bulk SU(3) theory, such  that the low-energy dynamics is that of 2+1
dimensional SU(2)$\times$U(1) theory, and probe the properties of 
the non-Abelian part, then the pattern should again be  largely
similar to what we   found in this paper. In other words, the domain
wall has to be   wider than the inverse of the bulk confinement scale
for   a 2+1 dimensional zero-mode to exist, but in general the
confinement  scale of the corresponding 
low-energy effective theory is smaller than that in the bulk
only by some  numerical factor roughly of order unity,   unless the
domain wall is significantly  wider
than the inverse of the bulk confinement scale. In the latter case 
the masses of localised bound states (or, say, the string tension
$\sqrt{\sigma}$)  are expected to scale as  $\sim g^2 \sim
g_4^2(\ell^{-1}) \ell^{-1}$, where $g_4^2(\bmu)$ is the renormalised
4d gauge coupling. For $1/\ell$ much below the 4d Yang-Mills 
confinement scale they are
thus of the same order of magnitude or larger than the lightest of
the higher modes, with masses $\sim 1/\ell$. 

The situation in the 3+1 dimensional case can perhaps be illustrated
by noting that,  in a way, the present mechanism is analogous to the
familiar dimensional  reduction of 4d Yang-Mills theory at a finite
temperature $T$, which occurs  only for $T$ above the deconfinement
phase transition temperature.  The role of $T$ is played  by
$\ell^{-1}$. The conflict arises because we now need a domain wall
wider than the bulk confinement scale,  corresponding to $T$ below
the deconfinement phase transition,  thus no dimensional reduction.

It would naturally be very interesting to understand whether  the
patterns observed in this paper change if we go to the physically
interesting  $4+n$ dimensional case, with $n \ge 1$.

\section*{Acknowledgements}

We thank D.~B\"odeker, P.~de Forcrand and M.~Shifman  for helpful
discussions. This work was partly supported by the RTN network {\em
Supersymmetry  and the Early Universe}, EU Contract No.\
HPRN-CT-2000-00152, by the Swiss Science Foundation,
and by the Academy of Finland,  grant No.~104382. H.M.\ thanks the
University of Lausanne for the generous {\em Bourse de
perfectionnement et de recherche}. 




\end{document}